\documentclass{kapproc}			

\usepackage{t1enc}

\usepackage{procps} 

\usepackage[dvips]{graphicx}

\upperandlowercase

\kluwerbib 

\def\spose#1{\hbox to 0pt{#1\hss}}
\def\simlt{\mathrel{\spose{\lower 3pt\hbox{$\mathchar"218$}}
     \raise 2.0pt\hbox{$\mathchar"13C$}}}
\def\simgt{\mathrel{\spose{\lower 3pt\hbox{$\mathchar"218$}}
     \raise 2.0pt\hbox{$\mathchar"13E$}}}

\begin{document}

\articletitle{Merger-Induced Starbursts}


\author{Fran\c{c}ois Schweizer}	
\affil{Carnegie Observatories, Pasadena, CA 91101, USA}
\email{schweizer@ociw.edu}

\begin{abstract}
Extragalactic starbursts induced by gravitational interactions can now be
studied from $z\approx 0$ to $\simgt$2.  The evidence that mergers of
gas-rich galaxies tend to trigger galaxy-wide starbursts is strong, both
statistically and in individual cases of major disk--disk mergers.
Star formation rates appear enhanced by factors of a few to $\sim$10$^3$
above normal.  Detailed studies of nearby mergers and ULIRGs suggest that
the main trigger for starbursts is the rapidly mounting pressure of the
ISM in extended shock regions, rather than high-velocity, 50\,--\,100
km~s$^{-1}$ cloud--cloud collisions.  Numerical simulations demonstrate
that in colliding galaxies the
star formation rate depends not only on the gas density,
but crucially also on energy dissipation in shocks.  An often overlooked
characteristic of merger-induced starbursts is that the spatial
distribution of the enhanced star formation extends over large scales
($\sim$10\,--\,20 kpc).  Thus, although most such starbursts do peak
near the galactic centers, young stellar populations pervade merger
remnants and explain why (1) age gradients in descendent galaxies are mild
and (2) resultant cluster systems are far-flung.   This review presents
an overview of interesting phenomena observed in galaxy-wide starbursts
and emphasizes that such events continue to accompany the birth of
elliptical galaxies to the present epoch.
\end{abstract}


\setcounter{page}{143}

\section{Introduction}
This brief review concentrates on three items.  First, I report on recent
progress in our understanding of the dynamical triggers at work in
merger-induced starbursts.  Then I address two issues that are often
ignored or misunderstood, yet are of fundamental importance to the subject:
the large spatial extent of merger-induced starbursts, and the implications
of such starbursts for the formation of elliptical galaxies at low and
high redshifts.

The basic reason why tidal interactions and mergers help fuel bursts of
star formation has been understood for over three decades.  Under the
headline {\em Stoking the Furnace?} Toomre \& Toomre (1972) wrote:
``Would not the violent mechanical agitation of a close tidal encounter---let
alone an actual merger---already tend to bring {\em deep} into a galaxy a
fairly {\em sudden} supply of fresh fuel in the form of interstellar
material?''  Subsequent numerical simulations that included gas have
fully corroborated this notion (e.g., Negroponte \& White 1983; Noguchi 1988;
Barnes \& Hernquist 1991).

Similarly, many observational studies have established beyond any
doubt that mergers invigorate star formation well beyond the
levels observed in quiescent disk galaxies.  As early as 1970, Shakhbazian
pointed out the presence of extraordinary stellar {\em ``superassociations''}
with luminosities of up to $M_V\approx -15.5$ in The Antennae. In a landmark
paper, Larson \& Tinsley (1978) showed that the $U\!BV$ colors of Arp's
peculiar galaxies can best be explained if tidal interactions engender
short, but intense bursts of star formation involving up to $\sim$\,5\% of
the total mass.  Infrared observations confirmed the notion of
{\em super}starbursts in major disk--disk mergers (Joseph \& Wright 1985).
Since then, a steady stream of papers from surveys (e.g., 2dF: Lambas et
al.\ 2003; SDSS: Nikolic et al.\ 2004) has continued to support and refine
this picture.  A nice twist was the discovery---fostered by {\em HST\,}'s
high resolution---that star clusters and, specifically, globular clusters
form in large numbers during galactic mergers (Schweizer 1987; Holtzman et
al.\ 1992; Whitmore et al.\ 1993).

\section{Gas Supply and Dynamical Triggers}

Even after $\sim$13 Gyr of evolution, many present-day galaxies still have
significant gas supplies available for star formation during interactions
and mergers.  The median gas fraction of neutral hydrogen alone, expressed
relative to the total baryonic mass, is 15\%, 10\%, and 4\% for dIrr, Sc,
and Sa galaxies, respectively (Roberts \& Haynes 1994).  Even more
impressive is the median fraction of {\em all} gas relative to the
dynamical mass, $M_{\rm H\,I\,+\,H_2}/M_{\rm dyn}\approx 25$\%, 15\%,
and 3\% for the same three types of galaxies (Young \& Scoville 1991).
Since even at high redshifts no galaxy can be more than 100\% gaseous,
the relatively high gas fractions of local Sc and later-type galaxies tell
us that there is less than one order-of-magnitude difference between the
fractional gas contents of many local disk galaxies and their high-$z$
counterparts.  Hence, the often-heard objection that mergers at
$z\approx 2$\,--\,5 were completely different from local mergers is weak, and
studying local mergers can, in fact, help us understand high-$z$ mergers.

Molecular gas masses observed in local mergers and distant quasars support
this point of view.  Locally, $M_{\rm H_2}$ ranges from
0.6$\times$10$^{10}\,M_{\odot}$ for an aging merger remnant like NGC 7252
through 1.5$\times$10$^{10}\,M_{\odot}$ for an ongoing merger like The
Antennae to $\sim$3$\times$10$^{10}\,M_{\odot}$ for extreme ULIRGs,
while $M_{\rm H_2} \approx 2\times10^{10}\,M_{\odot}$ in a QSO at
$z=2.56$ (Solomon et al.\ 2003) and also in one at $z=6.42$ (Walter et
al.\ 2003).

With gas amply available for star formation during mergers at
both low and high redshifts, what are the {\em dynamical triggers} for
merger-induced starbursts?

During disk-galaxy interactions gravitational torques arise between the
bars induced in the gas and among the stars.  Because the gaseous bar
leads, the gas experiences braking, which in turn leads to its infall.
The resulting pressure increase in the gas has long been understood to be
the root cause of interaction-induced starbursts (Noguchi 1988; Hernquist
1989; Barnes \& Hernquist 1991).

Although the vehemence of this pressure increase clearly depends on such
factors as the presence or absence of a central bulge (Mihos \& Hernquist
1996) and the encounter geometry (Barnes \& Hernquist 1996),
the small-scale details of the dynamical triggers have been less clear
until recently.

It now appears that {\em shocks} play a very major role, both in affecting
the spatial distribution of star formation (Barnes 2004) and in squeezing
giant molecular clouds (hereafter GMCs) into rapid star and cluster
formation (Jog \& Solomon 1992; Elmegreen \& Efremov 1997).  Questions
such as whether cloud--cloud collisions are important and what role
magnetic fields play are just beginning to be addressed by observers,
as detailed below.

\begin{figure}[t]
\includegraphics[height=5.4cm]{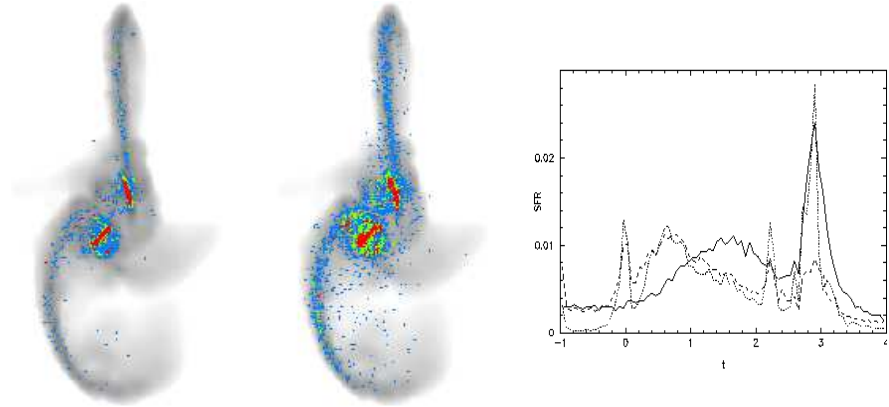}
\caption{
Simulations of star formation in The Mice for (left) density-dependent and
(middle) shock-induced star formation recipes; halftones mark old stars,
points mark star formation. (Right) Star formation rate vs.\ time for
density-dependent ({\em solid line}) and various shock-induced ({\em dashed \&
dotted}) star formation recipes (from Barnes 2004).
}
\end{figure}

Merger-induced shocks can be fierce.  In a simulation of two merging
equal-mass disks (Barnes \& Hernquist 1996), massive rings of dense gas
form around the
center of each galaxy and collide, during the third passage, head-on with
a relative velocity of $\sim$500 km s$^{-1}$!  This extreme final smash is
made possible by the rapid $\sim$90\% loss of orbital angular momentum that
the two gas rings experience within $\sim$1/4 disk-rotation period.

Even during milder encounters shock-induced star formation may dominate.
Because of the ubiquity of shocks in mergers involving gas, Barnes (2004)
proposes a new star formation recipe that, in addition to the local gas
density $\rho_{\rm gas}$, includes the local rate $\dot{u}$ of mechanical
heating due to shocks and $PdV$ work:\linebreak
\centerline{
      $ \dot{\rho_{\ast}} = C_{\ast} \cdot \rho^n_{\rm gas} \cdot
                            {\rm max}(\dot{u},0)^m\,. $
}
Assuming that energy dissipation balances the heating rate, setting
$m>0$ and $n=1$ yields purely {\em shock-induced} star formation, while
setting $m=0$ and $n>1$ yields {\em density-dependent} star formation
(with Schmidt's law as a special case).  Barnes compares simulations run for
these two limit cases of star formation with observations of The Mice (see
Fig.~1 here, and color figs.~3 \& 4 in his paper) and shows convincingly
that shock-induced star formation is {\em spatially more extended} and
{\em occurs earlier} during the merger, which is in significantly better
accord with the observations.

One long-standing question has been whether high-velocity cloud--cloud
collisions (50\,--\,100 km s$^{-1}$) contribute significantly to the
triggering of starbursts (Kumai, Basu, \& Fujimoto 1993) or not. To
address this issue, Whitmore et al.\ (in prep.) measured H$\alpha$ velocities
of the gas associated with young massive clusters in The Antennae, using
{\em HST}/STIS and positioning the 52${\tt ''}$ long slit of STIS along
different groups of clusters.  From many clusters in 7 regions the measured
cluster-to-cluster velocity dispersion is\,\ $<$10\,--\,12 km s$^{-1}$,
which argues {\em against} high-velocity cloud--cloud collisions as a major
trigger of starbursts.  Instead, the squeezing of GMCs by the general
pressure increase in the ISM (Jog \& Solomon 1992) appears favored.

The role of magnetic fields in triggering starbursts in mergers remains
unclear at present, but is beginning to be studied observationally.
Chy\.zy \& Beck (2004) have used the VLA to produce detailed maps of
radio total power and polarization in NGC 4038/39 (see Chy\.{z}y's
poster paper).  The derived mean total magnetic field of
$\sim$20\,$\mu$G is twice as strong as
in normal spirals and appears {\em tangled} in regions of enhanced star
formation. The field peaks at $\sim$30\,$\mu$G in the southern part
of the Overlap Region, suggesting strong compression where the star
formation rate is highest.  The crucial question to address over the
coming years is whether the enhanced magnetic field observed in mergers
merely traces compression, or whether it contributes to the triggering of
starbursts.

\section{Spatially Extended Starbursts}

Interaction-induced starbursts tend to be spatially extended
($\sim$10\,--\,20 kpc) for most of their duration.  Only relatively
late in a merger do they become strongly concentrated.

\begin{figure}[t]
\includegraphics[height=4.2cm]{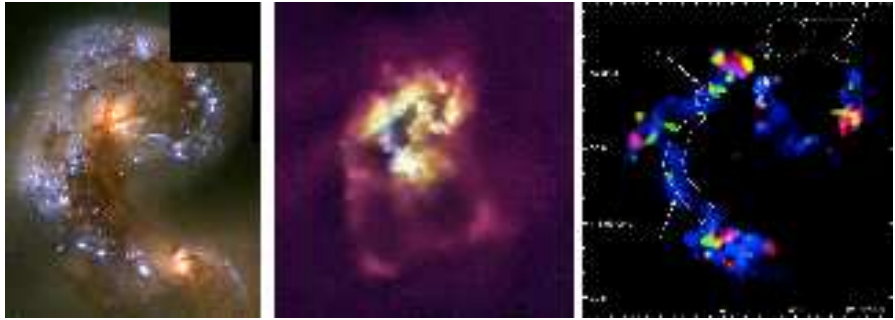}
\caption{
Extended starburst in NGC 4038/39, as imaged by (left) {\em HST}
($\sim$8$\,\times\,$11 kpc field of view) and (middle) {\em Chandra}. (Right)
Metallicity map for hot ISM, showing spotty chemical enrichment
(from Whitmore et al.\ 1999; Fabbiano et al.\ 2004; and Baldi et al.\
2005).
}
\end{figure}

For example, any good {\em HST, Spitzer,} or {\em Chandra} image of
NGC 4038/39 shows that enhanced star formation extends over a projected
area of $\sim$8$\times$11 kpc (Fig.~2).  In the optical, H$\alpha$ and
blue images are best at showing the extended nature of the starburst.
In the infrared, a {\em Spitzer}/IRAC image at 8$\,\mu$m emphasizes the
warm dust associated with star formation throughout the two disks,
glowing especially bright in the optically obscured disk contact
(``Overlap'') region (Wang et al.\ 2004). 
And in X-rays, a deep {\em Chandra} image displays
not only two disks filled with superbubbles of hot gas (typical
diameter $\sim$1.5~kpc, $T\approx$ 5$\times$10$^6$\,K, $M\approx
10^{5-6}M_{\odot}$), but also two giant, 10 kpc-size loops extending to
the south (Fig.\ 2, middle panel).  Their nature remains unclear (wind-blown?,
or tidal ejecta?).  These various images illustrate that
early in a merger the \mbox{extended starburst} heats the ISM in a chaotic
manner, rather than leading to well-directed bipolar superwinds.

An interesting consequence of the extended starburst in NGC 4038/39
is the {\em spotty chemical enrichment} of the hot ISM, observed for the
first time in any merger galaxy (Fabbiano et al.\ 2004; Baldi et al.\
2005).  The high $S/N$ ratio of the {\em Chandra} emission-line
spectra permits the determination of individual Fe, Ne, Mg, and Si
abundances in $\sim$20 regions.  Figure 2 (3rd panel, = color Fig.\ 3
in Fabbiano et al.) shows a metallicity map, with various shades of
gray marking individual elements.  The $\alpha$-elements are enhanced
by up to\,\ 20\,--\,25$\times$ solar and follow an enhancement pattern
distinctly different from Fe, as one would expect if they were recently
produced by SNe II. A question for future study is how such spotty
chemical enrichment may affect stars still to form.

Another important consequence of the large spatial extent of
merger-induced starbursts is that newly-formed stars decouple from the
inward-trending gas continuously and at many different radii.
This process differs sharply from the widely held misconception that
such starbursts occur mainly in the central kiloparsec, where they are
being fueled by infalling gas.
As a result of this extended star formation, radial age gradients in
merger remnants are weak (e.g., Schweizer 1998).
This is also the likely reason why in ellipticals age gradients
are near zero, and mean metallicity gradients are only $\sim$40\% per
decade in radius (Davies et al.\ 1993; Trager et al.\ 2000; Mehlert et
al.\ 2003).

\begin{figure}[t]
\hfill
\includegraphics[height=5.0cm]{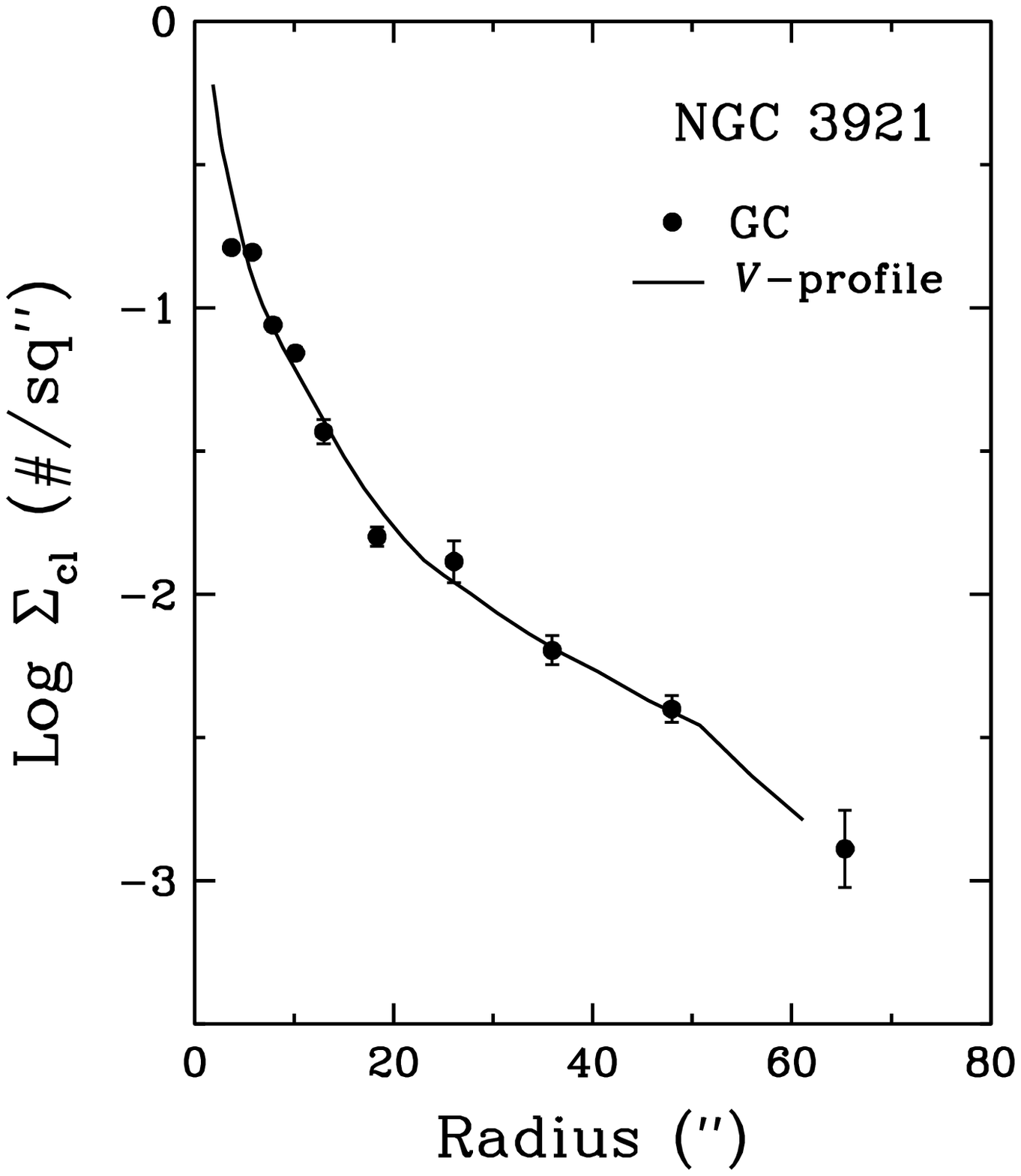}
\hfill
\includegraphics[height=5.0cm]{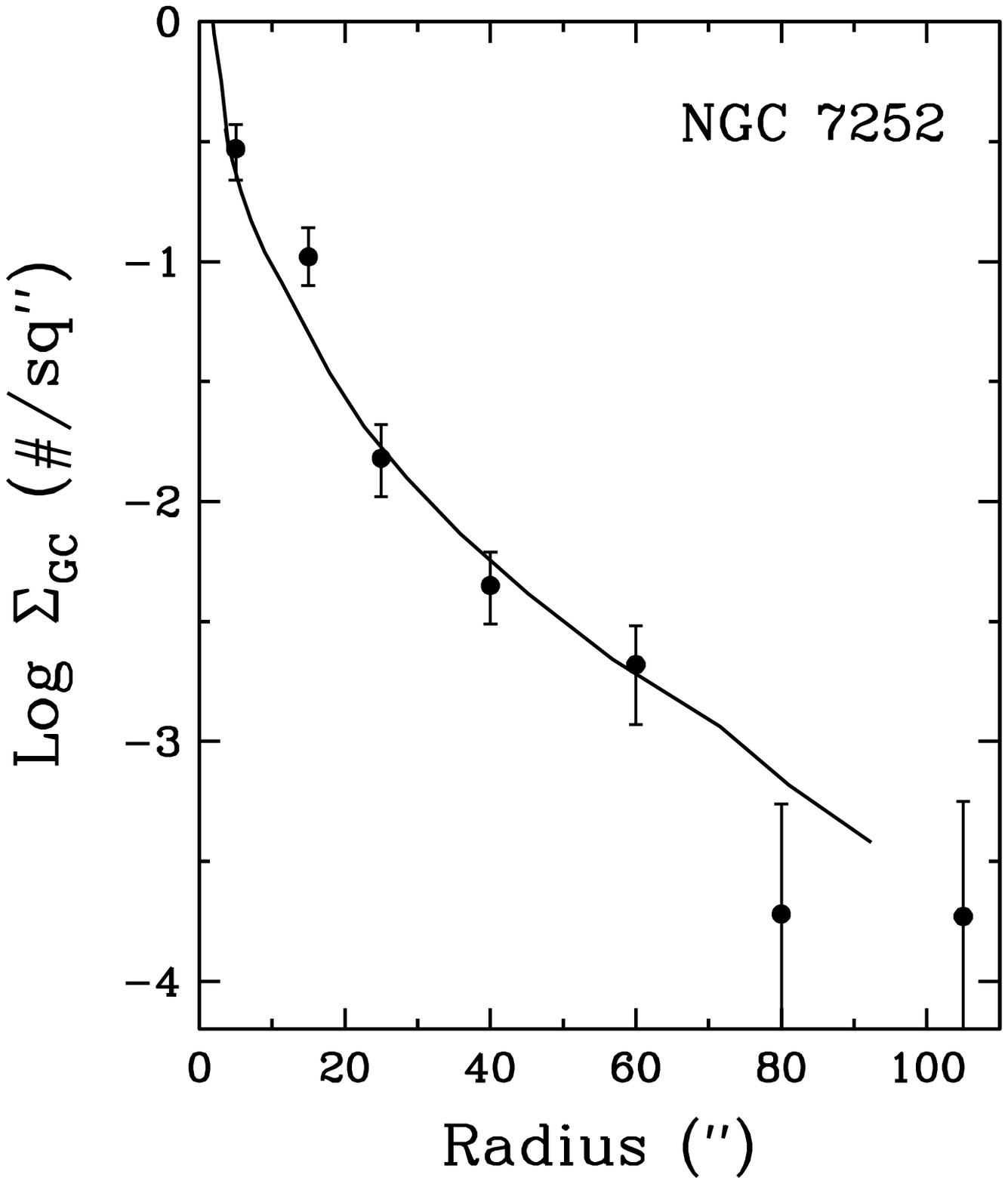}
\hfill
\caption{
Radial distributions of second-generation globular clusters ({\em data
points}) and background $V$-light ({\em lines}) in the merger remnants
NGC 3921 (Schweizer et al.\,1996) and NGC 7252 (Miller et al.\,1997).
The far-flung cluster distributions are remnant signatures of extended
starbursts.
}
\end{figure}

The strongest evidence linking extended starbursts to merger remnants
and ellipticals is the wide radial distribution of the resultant star
clusters.  In both remnants (Fig.~3) and Es,
second-generation metal-rich globular clusters track the underlying
light distribution of their host galaxies with surprising accuracy.
It is true that they tend to be more centrally distributed than
metal-poor globulars, but only by little.  Typically, half of them
lie within $R_{\rm eff}\approx 3$\,--\,5~kpc from the center.
This is consistent with some additional gaseous dissipation, but
completely inconsistent with nuclear-only ($\simlt$1~kpc) starbursts.
Hence wide-flung globular-cluster systems are signatures of ancient
extended starbursts.

\section{Cosmological Implications}

In 1972, Toomre \& Toomre put forth the bold hypothesis that most
giant ellipticals might be the remnants of major disk--disk mergers.
Toomre (1977) elaborated on this idea, proposing a sequence of 11
increasingly merged disk pairs and refining the argument that from
the current merger rate one could expect between 1/3 and all local
ellipticals to be remnants of ancient mergers.  Much evidence has
since accumulated to support this hypothesis.

Yet, beginning with the 1996 release of the {\em Hubble Deep Field}
data, a new generation of astronomers has begun to study galaxy
formation directly at high redshifts, often with remarkable success,
but too often also making claims about elliptical formation that run
afoul of the merger hypothesis and its strong supporting evidence
in the local universe.  For example, claims about (1) an
``E formation epoch'' ending around $z\approx 2$ and (2) constant
comoving space densities of ellipticals since then abound, but are
clearly mistaken.

Few astronomers would contest that disk--disk mergers are occurring
locally ($z\approx 0$) and forming remnants remarkably similar to
young ellipticals.
Evidence that some field Es contain intermediate-age stellar populations
is also increasingly being  accepted.
What remains controversial is how most older ellipticals formed, say
the majority that formed during the first half of the Hubble time
and now appear uniformly old, crammed as they are into a small, 0.3-dex 
logarithmic-age interval.
Did they form by major disk mergers as well, or did they form by a
process more akin to ``monolithic collapse''?

First, the similarities between recent, $\simlt$1 Gyr-old merger remnants
like NGC 3921 or NGC 7252 and giant Es (e.g., Toomre 1977; Schweizer
1982, 1996; Barnes 1998) are worth re-emphasizing.  The above two
remnants currently have luminosities of $\sim$2.8\,$L^{\ast}_V$ and will
still shine with $\sim$1.0\,$L^{\ast}_V$ after 10\,--\,12 Gyr of evolution.
They feature $r^{1/4}$-type light distributions, power-law cores, $U\!BV\!I$
color gradients, and velocity dispersions typical of Es.  Both also possess
many young, metal-rich halo globular clusters.  They show integrated
``E\,+\,A'' spectra indicative of $b$\,$\simgt$\,10\% starbursts
(Fritze-von Alvensleben \& Gerhardt 1994), as do many other similar young
merger remnants in the local universe (Zabludoff et al.\ 1996).
Hence, claiming that E formation ceased around $z$\,$\approx$\,2 is as
mistaken as would be any claim that star formation ceased then.
Local starbursts and merger remnants tell a different story.

Second, although the age distribution of local E and S0 galaxies is clearly
weighted toward old ages, it does show a tail of youngish galaxies, especially
in the field, with luminosity-weighted mean population ages of
$\sim$1.5\,--\,5 Gyr (Gonzalez 1993; Trager et al.\ 2000; Kuntschner et al.\
2002).  Hence, in the field E\,+\,S0 formation has clearly not ceased yet.

Third and to astronomers' surprise, massive disk galaxies not unlike the
Milky Way have been discovered at 1.4\,$\simlt$\,$z$\,$\simlt$\,3.0 (Labb\'e
et al.\ 2003) and thus {\em were available for major mergers at the epoch of
peak QSO formation}.  These galaxies seem to represent $\sim$half of all
galaxies with $L_V\!\geq 3L^{\ast}_V$ at those redshifts.  Complementing
such IR--optical observations, Genzel et al.\ (2003) have found a large
disk galaxy at $z$\,=\,2.8 whose rapidly rotating CO disk indicates a
dynamical mass of $\simgt$3$\times$10$^{11}M_{\odot}$.  Even more surprising
is a massive {\em old} disk galaxy at $z$\,=\,2.48 that shows a
pure exponential disk ($\alpha\approx 1.7$ kpc) and no bulge, has a
luminosity of $\sim$2$L^{\ast}_V$, and has not formed stars for the past
$\sim$2~Gyr (Stockton et al.\ 2004). This galaxy indicates that massive
Milky-Way-size disks were available for E formation through major mergers
even at $z>3$.

\begin{figure}[t]
\includegraphics[width=5.8cm,height=5.75cm]{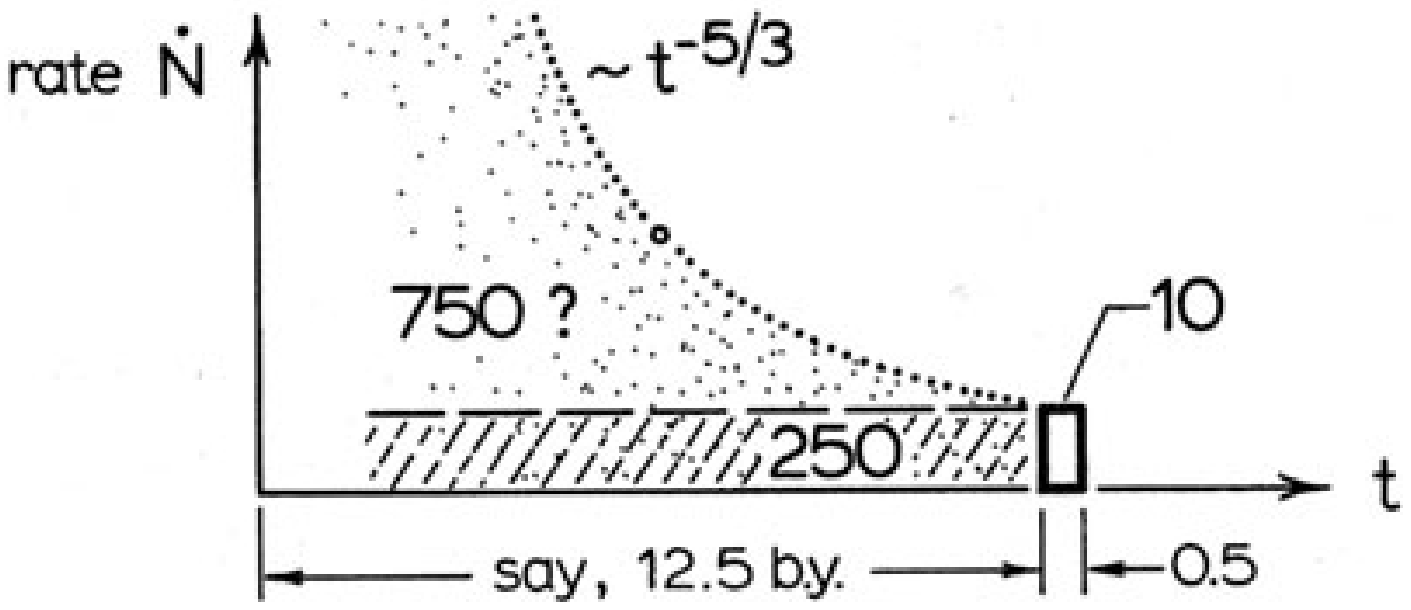}
\hfill
\includegraphics[width=5.7cm]{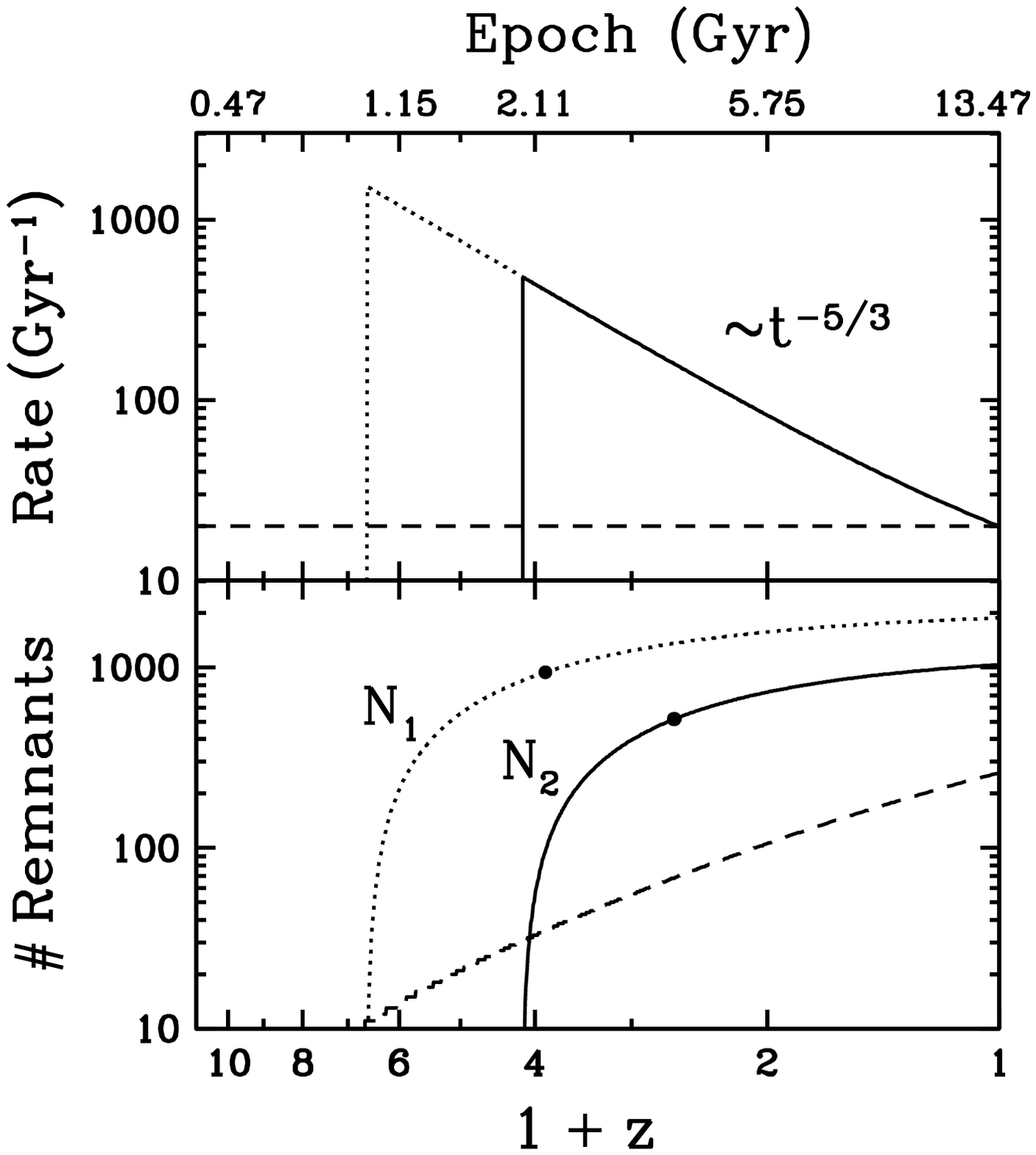}
\caption{
Merger rates and numbers of merger remnants as functions of cosmic
epoch. (Left) Toomre's (1977) original sketch and (right) a modern
version of it.  For details, see text.
}
\end{figure}

With this high-$z$ availability of disks and the above evidence that
disk mergers continue to form E-like remnants to the present epoch,
it is instructive to revisit Toomre's (1977) argument that most ellipticals
may be merger remnants.
Figure 4 shows, to the left, his original sketch of the merger rate
$\dot{N}(t)$ as a function of time $t$ and, to the right, a modern version
of it, in which I have plotted the rate and computed number of remnants vs.\
$(1+z)$.
From the $\sim$10 ongoing disk--disk mergers among $\sim$4000+ NGC galaxies
and their mean age of $\sim$0.5~Gyr, Toomre argued that there should
be {\em at least} 250 remnants among these NGC galaxies, had the rate
stayed constant, and more likely $\sim$750 remnants if the merger rate
declined as $t^{-5/3}$ (assuming a flat distribution of binding energies
for binary galaxies).
The latter number being close to the number of Es in the catalog, he
suggested that {\it most\,} such galaxies may be old merger remnants.

The top panel of the modern diagram shows the same rate,\,\ 
$\dot{N}\propto t^{-5/3}$,\,\ plotted vs.\ $(1+z)$ assuming that major
disk merging began 1~Gyr ({\em dotted lines}) or 2~Gyr ({\em solid})
after the big bang, with the corresponding numbers of remnants labeled
$N_1$ and $N_2$ in the bottom panel.
Dashed lines mark the case of constant $\dot{N}(t)$ for comparison, and
epochs for the standard $\lambda$CDM cosmology are given at the top.
Notice that major disk mergers beginning at 1~Gyr ($z\approx 5.6$) would
produce more remnants than needed to explain all Es among present-day
NGC galaxies, while such mergers beginning at 2~Gyr ($z=3.15$) would produce
just about the right number.  Interestingly, half of the $N_2$ remnants
would already have formed at $z=1.64$ ({\em dot} on $N_2$ curve), which may
explain why observers are having a hard time deciding whether the comoving
space density of Es changes from $z\approx 1.5$ to 0 or not.

In summary, with massive disk galaxies already present at $z\simgt 3$,
major disk mergers must have contributed to a growing population of
elliptical galaxies ever since.  Like star formation, E formation through
major mergers is an ongoing process in which gaseous dissipation and
starbursts play crucial roles.

\paragraph{Acknowledgments}
I thank A.\ Baldi, J.E.\ Barnes, G.\ Fabbiano, A.\ Toomre, and B.C.\ Whitmore
for permission to reproduce figures, and gratefully acknowledge support
from the NSF through Grant AST--02\,05994.



\begin{chapthebibliography}{1}

\bibitem[]{} Baldi, A., et al. 2005, ApJ, in press (astro-ph/0410192)

\bibitem[]{} Barnes, J.E.  1998, in Galaxies: Interactions and Induced Star
Formation, ed.\ D.\ Friedli, L.\ Martinet, \& D. Pfenniger (Berlin:
Springer), 275

\bibitem[Barnes(2004)]{barn04} Barnes, J.E. 2004, MNRAS, 350, 798

\bibitem[]{} Barnes, J.E., \& Hernquist, L.E. 1991, ApJ, 370, L65

\bibitem[]{} Barnes, J.E., \& Hernquist, L. 1996, ApJ, 471, 115

\bibitem[]{} Chy\.{z}y, K.T., \& Beck, R. 2004, A\&A, 417, 541

\bibitem[]{} Davies, R.L., Sadler, E.M., \& Peletier, R.F. 1993, MNRAS,
  262, 650

\bibitem[Elmegreen \& Efremov(1997)]{elme97} Elmegreen, B.G., \& Efremov,
   Y.N. 1997, ApJ, 480, 235

\bibitem[]{} Fabbiano, G., et al. 2004, ApJ, 605, L21

\bibitem[]{} Fritze-von Alvensleben, U., \& Gerhard, O.E. 1994, A\&A, 285, 775

\bibitem[]{} Genzel, R., et al. 2003, ApJ, 584, 633

\bibitem[Gonz\'alez(1993)]{gonz93} Gonz\'alez, J.J. 1993, Ph.D. thesis,
   UC Santa Cruz

\bibitem[]{} Hernquist, L. 1989, Nature, 340, 687

\bibitem[Holtzman et al.(1992)]{holt92} Holtzman, J.A., et al. 1992, AJ,
   103, 691

\bibitem[Jog \& Solomon(1992)]{jog92} Jog, C.J., \& Solomon, P.M. 1992,
   ApJ, 387, 152

\bibitem[]{} Joseph, R.D., \& Wright, G.S. 1985, MNRAS, 214, 87

\bibitem[]{} Kumai, Y., Basu, B., \& Fujimoto, M. 1993, ApJ, 404, 144

\bibitem[Kuntschner et al.(2002)]{kunt02} Kuntschner, H., Smith, R.J.,
   Colless, M., Davies, R.L., Kaldare, R., \& Vazdekis, A. 2002, MNRAS,
   337, 172

\bibitem[]{} Labb\'e, I., et al. 2003, ApJ, 591, L95

\bibitem[]{} Lambas, D.G., Tissera, P.B., Alonso, M.S., Coldwell, G. 2003,
   MNRAS, 346, 1189

\bibitem[]{} Larson, R.B., \& Tinsley, B.M. 1978, ApJ, 219, 46

\bibitem[]{} Mehlert, D., Thomas, D., Saglia, R.P., Bender, R., \&
   Wegner, G. 2003, A\&A, 407, 423

\bibitem[]{} Mihos, J.C., \& Hernquist, L. 1996, ApJ, 464, 641

\bibitem[]{mill97}Miller, B.W., Whitmore, B.C., Schweizer, F., \& Fall, S.M.
   1997, AJ, 114, 2381

\bibitem[]{} Negroponte, J., \& White, S.D.M.  1983, MNRAS, 205, 1009

\bibitem[]{} Nikolic, B., Cullen, H., \& Alexander, P. 2004, MNRAS, in press
   (astro-ph/0407289)

\bibitem[]{} Noguchi, M. 1988, A\&A, 203, 259

\bibitem[]{} Roberts, M.S., \& Haynes, M.P. 1994, ARA\&A, 32, 115

\bibitem[]{} Schweizer, F. 1982, ApJ, 252, 455

\bibitem[]{schw87}Schweizer, F. 1987, in Nearly Normal Galaxies, ed.\ S.M.\
   Faber (New York: Springer), p.~18

\bibitem[]{} Schweizer, F. 1996, AJ, 111, 109

\bibitem[]{} Schweizer, F. 1998, in Galaxies: Interactions and Induced Star
Formation, ed.\ D.\ Friedli, L.\ Martinet, \& D. Pfenniger (Berlin:
Springer), 105               

\bibitem[Schweizer et al.(1996)]{schw96b} Schweizer, F., Miller, B.W.,
   Whitmore, B.C., \& Fall, S.M. 1996, AJ, 112, 1839

\bibitem[]{} Shakhbazian, R.K. 1970, Afz, 6, 367 (= Astrophysics, 6, 195)

\bibitem[]{} Solomon, P., Vanden Bout, P., Carilli, C., \& Guelin, M. 2003,
   Nature, 426, 636

\bibitem[]{} Stockton, A., Canalizo, G., \& Maihara, T. 2004, ApJ, 605, 37

\bibitem[Toomre \& Toomre(1972)]{tt72} Toomre, A., \& Toomre, J. 1972, ApJ,
   178, 623

\bibitem[Toomre(1977)]{t77}Toomre, A. 1977, in The Evolution of Galaxies and
   Stellar Populations, ed.\ B.M.\ Tinsley \& R.B.\ Larson (New Haven: Yale
   Univ.\ Obs.), 401

\bibitem[]{} Trager, S.C., Faber, S.M., Worthey, G., \& Gonz\'alez, J.J. 2000,
   AJ, 119, 1645

\bibitem[]{} Walter, F., et al. 2003, Nature, 424, 406

\bibitem[]{} Wang, Z., et al. 2004, ApJS, 154, 193

\bibitem[Whitmore et al.(1993)]{whit93} Whitmore, B.C., Schweizer, F.,
   Leitherer, C., Borne, K., \& Robert, C. 1993, AJ, 106, 1354

\bibitem[Whitmore et al.(1999)]{whit99} Whitmore, B.C., Zhang, Q., Leitherer,
   C., Fall, S.M., Schweizer, F., \& Miller, B.W. 1999, AJ, 118, 1551

\bibitem[]{} Young, J.S., \& Scoville, N.Z. 1991, ARA\&A, 29, 581

\bibitem[]{} Zabludoff, A., et al. 1996, ApJ, 466, 104

\end{chapthebibliography}
\end{document}